\documentclass[aps,prb,reprint,a4paper,superscriptaddress,showpacs,floatfix]{revtex4-1}
\usepackage{bm}
\usepackage{amsmath,amssymb}
\usepackage{graphicx}
\usepackage{natbib}
\usepackage{color}

\begin{document}

\title{Magnetic spectrum of the two-dimensional antiferromagnet La$_2$CoO$_4$\\ studied by inelastic neutron scattering}

\author{P. Babkevich}
\email[]{peter.babkevich@physics.ox.ac.uk}
\affiliation{Department of Physics, Oxford University, Oxford, OX1 3PU, United Kingdom}
\affiliation{Laboratory for Neutron Scattering, Paul Scherrer Institut, CH-5232 Villigen PSI, Switzerland}
\author{D. Prabhakaran}
\affiliation{Department of Physics, Oxford University, Oxford, OX1 3PU, United Kingdom}
\author{C. D. Frost}
\affiliation{ISIS Facility, Rutherford Appleton Laboratory, Didcot, Oxon., OX11 0QX, United Kingdom}
\author{A. T. Boothroyd}
\affiliation{Department of Physics, Oxford University, Oxford, OX1 3PU, United Kingdom}

\date{\today}

\begin{abstract}
We report measurements of the magnetic excitation spectrum of the layered antiferromagnet La$_2$CoO$_4$ by time-of-flight neutron inelastic scattering. In the energy range probed in our experiments (0--250\,meV) the magnetic spectrum consists of spin-wave modes with strong in-plane dispersion extending up to 60\,meV, and a nearly dispersionless peak at 190\,meV. The spin-wave modes exhibit a small ($\sim 1$\,meV) dispersion along the magnetic zone boundary. We show that the magnetic spectrum can be described very well by a model of a Heisenberg antiferromagnet that includes the full spin and orbital degrees of freedom of Co$^{2+}$ in an axially-distorted crystal field. The collective magnetic dynamics are found to be controlled by dominant nearest-neighbour exchange interactions, strong XY-like single-ion anisotropy and a substantial unquenched orbital angular momentum.
\end{abstract}


\maketitle

\section{Introduction}
\label{sec:intro}

The evolution from antiferromagnetism to high-temperature superconductivity with carrier doping of the layered copper oxides\cite{bednorz-1986} has inspired a vast literature on the electronic, structural, dynamical, and chemical properties of related materials. It has become clear from these studies that superconductivity and commensurate antiferromagnetic order are only two out of many different competing ordering tendencies found in systems of strongly interacting electrons.

Among other forms of order found in doped Mott insulators are nematic phases characterised by unidirectional density-wave states involving combined charge and spin order. Such `striped' phases were first identified many years ago in hole-doped (La,Nd)$_2$CuO$_4$ (Ref.~\onlinecite{tranquada-nature-1995}) and La$_2$NiO$_4$ (Refs.~\onlinecite{chen-prl-1993,tranquada-prl-1994,wochner-prb-1998, yoshizawa-prb-2000}), but their significance for high-temperature superconductivity has been the subject of a continuing debate. Although much of the focus has been on the cuprates, the nickelates have contributed to this debate on account of their relatively well correlated and stable stripe order which is amenable to experimental investigation. One drawback, however, is that holes localised on Ni$^{3+}$ ions in hole-doped La$_2$NiO$_4$ carry a spin which can interact magnetically both with other spins in the charge stripes and with the surrounding antiferromagnetic matrix of Ni$^{2+}$. The influence of these interacting magnetic degrees of freedom on the properties of stripes in nickelates has yet to be fully evaluated, but spin correlations associated with both Ni sites have been observed\cite{boothroyd-prl-2003} and there remain some unexplained features in the spin excitation spectra.\cite{boothroyd-prb-2003}

Recently, evidence has been presented for the existence of stripe phases in the layered cobaltate system La$_{2-x}$Sr$_x$CoO$_4$ (Ref.~\onlinecite{cwik-prl-2009}), which is isostructural with hole-doped La$_2$CuO$_4$ and La$_2$NiO$_4$. Neutron diffraction measurements on half-doped ($x=0.5$) La$_2$CoO$_4$ show clear evidence for a charge ordering of Co$^{2+}$ and Co$^{3+}$ ions in a checkerboard pattern below $T_{\rm co} \approx 825$\,K, with an antiferromagnetic ordering transition at a much lower temperature $T_{\rm N} \sim 60$\,K.\cite{zaliznyak-prl-2000,savici-prb-2007,horigane-jpsj-2007} The antiferromagnetic order in La$_{2-x}$Sr$_x$CoO$_4$ in the range $0.3< x < 0.6$ was observed to be periodically modulated with the modulation wavevector nearly proportional to $x$,\cite{cwik-prl-2009} a characteristic of the stripe phase found in La$_{2-x}$Sr$_x$NiO$_4$. The Sr-doped cobaltate system has an advantage over the nickelates in that for the compositions in which stripe-like magnetic correlations occur there is evidence to suggest that the Co$^{3+}$ ions adopt the low spin ($S = 0$) state at low temperatures and are therefore not magnetically active.\cite{hollmann-journphys-2008,chang-prl-2009,helme-prb-2009}  Hence, the layered cobaltates offer the chance to investigate the fundamental interactions and excited states of an ordered stripe phase in which the doped holes do not possess low-energy spin degrees of freedom.

Attempts to understand the electronic phases in La$_{2-x}$Sr$_x$CoO$_4$ will require some basic knowledge of the parent antiferromagnet La$_2$CoO$_4$. Although the crystal structure and magnetic order of La$_2$CoO$_4$ have been studied in detail,\cite{yamada-prb-1989} no measurements of the magnetic excitation spectrum have been published before now.

In this paper we present high-resolution measurements of the magnetic spectrum of La$_2$CoO$_{4+\delta}$ $(\delta \approx 0)$ by neutron inelastic scattering. The methodology relates closely to that employed in a recent investigation carried out on the half-doped cobaltate La$_{1.5}$Sr$_{0.5}$CoO$_4$ (Ref.~\onlinecite{helme-prb-2009}). The data presented here on La$_2$CoO$_{4}$ extend up to 250\,meV in energy and reveal spin-wave-like excitations with a bandwidth of 60\,meV. At much higher energies, there are excitations of a localised character. A good description of the data is achieved with a spin-wave model for a quasi-two-dimensional antiferromagnet that includes the full spin and orbital degrees of freedom of the Co$^{2+}$ ions. The results show that La$_2$CoO$_4$ has dominant nearest-neighbour exchange interactions, although a weak dispersion along the zone boundary indicates that more distant interactions or non-linear terms in the Hamiltonian are not negligible. The anisotropy is strongly XY-like but there is also a weak in-plane anisotropy.

\section{Crystal and magnetic structure}
\label{sec:structure}
\begin{figure}
\includegraphics[width=0.9\columnwidth]{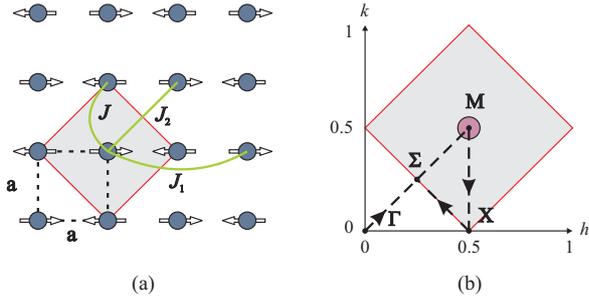}
\caption{  (a) In-plane magnetic structure of La$_2$CoO$_4$. The dashed square shows the conventional $I4/mmm$ unit cell of the HTT phase, and the filled square represents the magnetic unit cell, which coincides with the $\sqrt{2} \times \sqrt{2}$ chemical unit cell of the LTT phase. The exchange interactions used to model the magnetic spectrum are indicated. (b) Diagram of the reciprocal space lattice corresponding to the $I4/mmm$ cell. The filled square indicates the magnetic Brillouin zone centered on $(0.5,0.5)$. The dashed lines show the path through reciprocal space along high-symmetry directions used for detailed analysis of the magnetic excitation spectrum --- see Fig.~\ref{fig:calcdisp}.
\label{fig:spins}}
\end{figure}

In common with the stoichiometric La$_2$CuO$_4$ and La$_2$NiO$_4$ compounds, La$_2$CoO$_4$ exhibits three different structural phases:\cite{yamada-prb-1989} (i) $T>T_1$ high-temperature tetragonal (HTT), space group $I4/mmm$; (ii) $T_2 < T < T_1$  low-temperature orthorhombic (LTO), space group {\it Cmca}; (iii) $T<T_2$ low-temperature tetragonal (LTT), space group $P$4$_2$/$ncm$. The structural transition temperatures for La$_2$CoO$_4$ are $T_1 \approx 900$\,K (Ref.~\onlinecite{cwik-prl-2009}) and $T_2 =120-135$\,K (Ref.~\onlinecite{yamada-prb-1989}). The latter is reported to be first order. Throughout this paper we shall use the conventional $I4/mmm$ unit cell as a basis for the reciprocal lattice. The low temperature lattice constants referred to this cell are $a = b = 3.91$\,{\AA} and $c = 12.6$\,{\AA}. The true LTT unit cell has in-plane dimensions which are $\sqrt{2} \times \sqrt{2}$ larger than those of the $I4/mmm$ pseudo-cell --- see Fig.~\ref{fig:spins}(a).

The transition to magnetic order occurs at $T_{\rm N} \approx 275$\,K, and a magnetic reorientation occurs at $T_2$ coincident with the LTO--LTT structural transition. The antiferromagnetic structure has an ordering wavevector ${\bf q}_{\rm m} = (0.5,0.5,0)$, with ordered moments lying in the CoO$_2$ plane. Assuming collinear order, the difference between the magnetic structures in the LTT and LTO phases is that in the LTT phase the moments are perpendicular to ${\bf q}_{\rm m}$ whereas in the LTO phase they are parallel to ${\bf q}_{\rm m}$. The distinction between these structures depends on the relationship between adjacent layers.  Another possibility is that the structure is collinear within the layers but the moment direction rotates by $\pm 90^{\circ}$ from one layer to the next.\cite{yamada-prb-1989} In the absence of inter-layer coupling all these structures have the same energy. As we did not observe any evidence in the excitation spectrum for inter-layer coupling we will treat the magnetic order as two-dimensional. Figure~\ref{fig:spins}(a) shows the in-plane magnetic order with the moments arbitrarily chosen to point along the horizontal axis.

\section{Experimental Details}
\label{sec:expt}
% sample growth details
A single-crystal sample of La$_2$CoO$_4$ with a mass of approximately 5\,g was grown in Oxford by the optical floating-zone method. Polycrystalline La$_2$CoO$_4$ was prepared from high-purity ($>$$99.99$\%) La$_2$O$_3$ and Co$_3$O$_4$ by solid-state reaction.  Stoichiometric amounts of the oxides were mixed and reacted at 1050$^{\circ}$C for 48 hours under a flowing atmosphere of CO/CO$_2$ mixed in the ratio 1:10.  A reducing atmosphere is needed to avoid the formation of LaCoO$_3$. The powder was re-ground and sintered at 1100$^{\circ}$C in a flow of argon for a further 48 hours. No impurity phases could be detected in the product by x-ray powder diffraction.  The La$_2$CoO$_4$ powder was isostatically pressed into rods of diameter 12\,mm and length 120\,mm.  The rods were sintered in an argon atmosphere at 1250$^{\circ}$C for 24 hours.  Crystal growth was carried out in a four-mirror image furnace (Crystal Systems Corporation) in flowing argon at a growth speed of approximately 2\,mm\,hr$^{-1}$ with counter-rotation of the feed and seed rods at 25\,rpm.

Crystals grown by this method contain an excess of oxygen. To achieve stoichiometry the as-grown crystal was annealed at 850$^{\circ}$C for 72 hours in flowing CO/CO$_2$ (1:10 ratio). A fragment of the annealed crystal was ground to a powder and subjected to a thermogravimetric analysis. From the measured weight loss we determined the oxygen nonstoichiometry to be $\delta = -0.03 \pm 0.02$. This suggests that the crystal is close to the ideal stoichiometry, if anything slightly oxygen-deficient.

% Magnetization characterization
Magnetisation measurements were performed with a superconducting quantum interference device (SQUID) magnetometer (Quantum Design) on a small crystal cut from the same rod as the neutron scattering crystal. Measurements were made by the dc method with a measuring field of strength 1000\,Oe ($\mu_0 H = 0.1$\,T) applied along the crystallographic $c$ axis.  Zero-field-cooled (ZFC) data were recorded on warming after the sample had been initially cooled from 350\,K in zero applied field, and field-cooled (FC) data were recorded while cooling the sample from 350\,K in the measuring field.

% MAPS description
Unpolarised-neutron inelastic scattering measurements were performed on the direct-geometry chopper spectrometer MAPS at the ISIS facility.\cite{perring-maps-2004} Neutron time-of-flight instruments with large position sensitive detector arrays such as MAPS allow sampling of vast regions of $({\bf Q}, \omega)$ space simultaneously, where $\bf Q$ and $\hbar\omega$ are, respectively, the wavevector and energy transferred from the neutron to the sample. This is very advantageous in studies where the excitation spectrum is required throughout the Brillouin zone.

% Introduction to MAPS data
In preparation for the neutron measurements the La$_2$CoO$_4$ crystal was sealed in a thin-walled aluminium can containing helium exchange gas and aligned with the $c$ axis parallel to the direction of the incident-neutron beam. Cooling was provided by a closed-cycle refrigerator. Data were collected with incident-neutron energies of 51, 86, 111, 152 and 303\,meV. The energy resolution was typically 5\% of the incident energy (full width at half maximum) at zero energy transfer, decreasing slightly with increasing energy transfer. Under the chosen experimental conditions the wavevector resolution is largely determined by the divergence of the incident-neutron beam which is approximately 0.5$^{\circ}$.  Spectra from La$_2$CoO$_4$ were recorded at several temperatures between 6 and 300\,K. Separate measurements of a standard vanadium sample were made at each incident energy to normalise the spectra and place them on an absolute intensity scale.

For presentation and analysis, the neutron data were transformed from raw time-of-flight spectra into an intensity map as a function of $\bf Q$ and $\hbar\omega$. With a fixed sample orientation, only three out of the four components of $({\bf Q}, \omega)$ are independent. We chose the two in-plane wavevector components ($Q_x,Q_y) = (h, k)\times2\pi/a$ and energy as the independent variables, which means that the out-of-plane wavevector $Q_z = l\times2\pi/c$ varies implicitly with energy transfer. For a two-dimensional scattering system, however, there is no dispersion in the out-of-plane direction and the gradual variation of scattering intensity with $Q_z$ can be included in a model (and was done so in this work). The justification for treating La$_2$CoO$_4$ as a two-dimensional magnetic system is that the magnetic spectra show no discernible periodic modulation in intensity with $Q_z$ (i.e., with $\hbar\omega$).

In order to quantify the magnetic dispersion we performed a series of constant-energy and constant-wavevector cuts through the data volume along high-symmetry directions using the \textsc{MSLICE} software.\cite{coldea-mslice} Before performing these cuts, data at symmetry-equivalent wavevectors were averaged to improve the signal.

\section{Results}

\begin{figure}
\includegraphics[width=0.8\columnwidth]{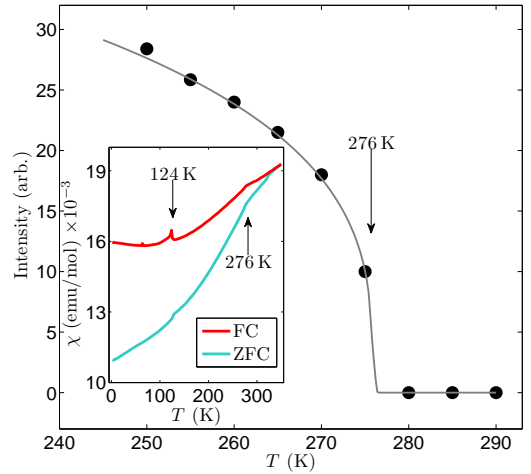}
\caption{  Main figure: temperature dependence of the $(1.5,1.5,1)$ magnetic Bragg peak of La$_2$CoO$_4$ showing the onset of magnetic order at $T_{\rm N} \approx 276$\,K. The solid line shows a power law fit to data with a Gaussian distribution of N\'{e}el temperatures. Inset: field-cooled (FC) and zero-field-cooled (ZFC) magnetic susceptibility of La$_2$CoO$_4$ recorded with a measuring field of 1000\,Oe applied parallel to the $c$ axis. The magnetic ordering transition and the LTO--LTT structural transition are indicated by arrows.
\label{fig:tdept}}
\end{figure}

\begin{figure*}
\includegraphics[width=0.8\textwidth,bb=0 413 592 798] {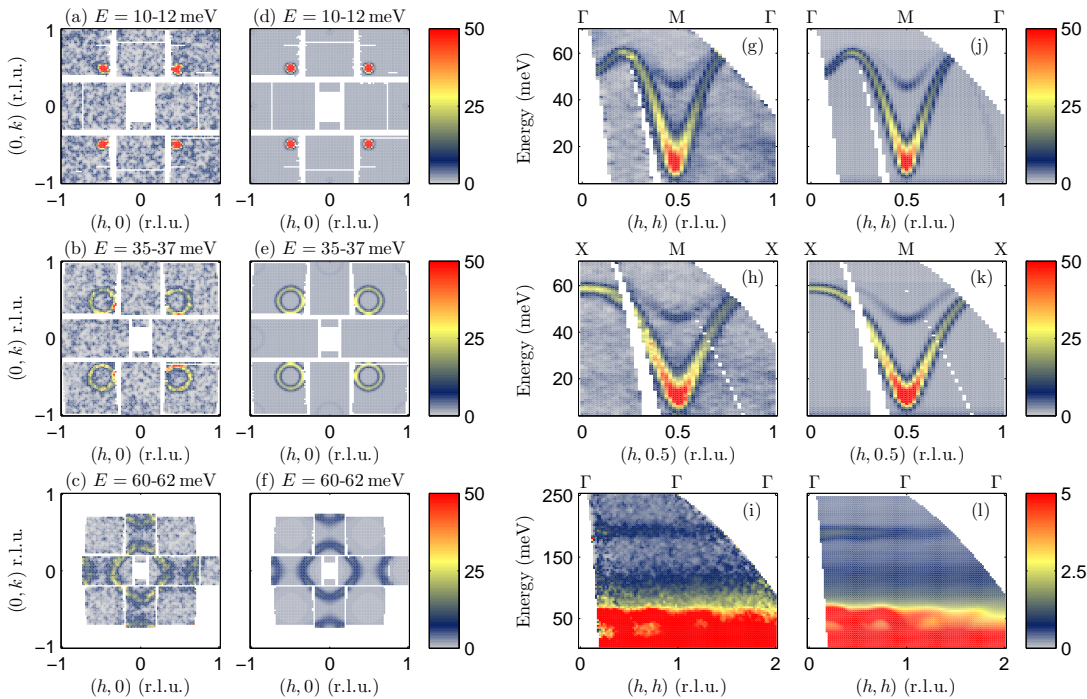}
\caption{  Measured and simulated magnetic spectra of La$_2$CoO$_4$ at 6\,K. Panels (a)--(c) show intensity maps averaged over 2\,meV energy ranges centered on different energies as indicated, with the corresponding calculated spectra shown in (d)--(f). The magnetic dispersion along two high-symmetry directions is displayed in (g)--(i), with corresponding simulations in (j)--(l). Data in (a)--(h) were measured with an incident neutron energy $E_{\rm i} = 86$\,meV, while (i) was measured with $E_{\rm i} = 303$\,meV. The units of intensity indicated by the colourbars are mb\,sr$^{-1}$\,meV$^{-1}$\,f.u.$^{-1}$. The simulated spectra are calculated from the spin-orbital spin-wave model described in the text.
\label{fig:MAPS_ExcitonQ}}
\end{figure*}

\begin{figure}[h!]
\includegraphics[width=\columnwidth,bb=15 221 590 600]    {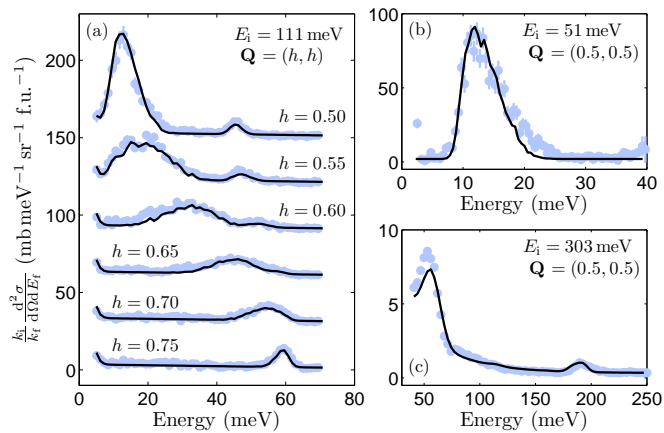}
\caption{  Constant-$\bf{Q}$ cuts taken through the measured and simulated spectra. Panel (a) shows a series of cuts following the dispersion from the zone center $(0.5,0.5)$ to the antiferromagnetic zone boundary at $(0.75,0.75)$ measured with incident energy 111\,meV. Panel (b) shows the spin gap of $\sim$10\,meV at $(0.5,0.5)$ measured with an incident energy of 51\,meV which gave an improved energy resolution. Panel (c) shows the measured and simulated excitation mode at $\sim190$\,meV. These data were obtained with $E_{\rm i}= 303$\,meV.
\label{fig:cuts}}
\end{figure}

% sample characterisation using SQUID and MAPS
The temperature dependence of the FC and ZFC susceptibility ($\chi=M/H$) is shown in Fig.~\ref{fig:tdept} (inset). Both curves show a change in slope at approximately $276$\,K consistent with the antiferromagnetic transition, and sharp anomalies at $124$\,K close to the temperature $T_2$ at which the LTO--LTT structural transition is expected. The onset of antiferromagnetism at $T_{\rm N} \approx 276$\,K is confirmed by the temperature dependence of the neutron diffraction intensity recorded at the magnetic Bragg peak position $(1.5,1.5,1)$, shown in Fig.~\ref{fig:tdept}. To estimate $T_{\rm N}$ we fitted the data to a power law  $I\propto ( 1 - T/T_{\rm N})^{2\beta}$ , assuming a Gaussian distribution of N\'{e}el temperatures about the mean value $\langle T_{\rm N}\rangle$ with standard deviation of $\sigma_T$. This function was found to give a good description of diffraction data near $T_{\rm N}$ in Ref.~\onlinecite{yamada-prb-1989}. The parameters obtained from our data were $\langle T_{\rm N} \rangle = 275.5(5)$\,K, $\sigma_T = 1(1)$\,K, and $\beta = 0.15(1)$. The transition temperatures measured on our sample are consistent with previously reported values of $T_{\rm N} = 275$\,K and $T_2 \approx 135$\,K for a nominally stoichiometric crystal.\cite{yamada-prb-1989} Although there is some discussion in the literature about the precise composition of La$_2$CoO$_4$ prepared under different conditions,\cite{lewandowski-JAmCeramSoc-1986,mohan-MRB-1988,yamada-prb-1989,kajitani-jpsj-1990} we can at the very least be confident that our crystal is close in composition to the one used in Ref.~\onlinecite{yamada-prb-1989}.

We note that the FC and ZFC susceptibility curves separate below 350\,K, which is not expected in the paramagnetic phase. This indicates that the sample contains a small amount of ferromagnetic impurity. The FC--ZFC separation was not observed in the as-grown crystal. The most probable explanation is that a tiny amount of elemental Co was formed during the CO/CO$_2$ annealing step. This is consistent with the slight oxygen deficiency found from the thermogravimetric analysis. As there is no unexplained secondary signal in our neutron scattering spectra this impurity must be present in very small quantities so is of no consequence to our neutron results, but it does mean that the susceptibility curves shown in Fig.~\ref{fig:tdept} contain a background signal in additional to the signal from pure La$_2$CoO$_4$.

% Qualitative description of MAPS data
We now turn to the neutron scattering spectra. Figure~\ref{fig:MAPS_ExcitonQ} provides an overview of the data collected at 6\,K. Panels (a)--(c) are constant-energy slices at three different energies, and panels (g)--(i) are energy--$\bf{Q}$ slices to illustrate the magnetic dispersion. The spectrum is dominated by a spin-wave-like conical dispersion which rises from the in-plane antiferromagnetic ordering wavevector ${\bf q}_{\rm m} = (0.5,0.5)$ and equivalent positions [the M-points of the square-lattice Brillouin zone --- see Fig.~\ref{fig:spins}(b)]. This mode has a gap of approximately 10\,meV at the M-point and rises to a maximum energy of 60\,meV at the $\Sigma$-point on the Brillouin zone boundary. A much weaker branch, displaying an upwards dispersion with a minimum energy at M of 46\,meV, corresponds to the first mode translated by ${\bf q}_{\rm m}$. The large splitting of the modes at M shows that the anisotropy is strongly XY-like. The lower and upper modes correspond to in-plane and out-of-plane fluctuations, respectively. Figure~\ref{fig:MAPS_ExcitonQ}(i) shows data up to the maximum energy explored in our experiment. This reveals only one other significant feature --- a band of scattering in a narrow range of energies close to 190\,meV. We note that scattering from phonons is much weaker than magnetic scattering in the range of ${\bf Q}$ studied.

To give a different impression of the data we present in Fig.~\ref{fig:cuts} examples of constant-$\bf{Q}$ cuts taken through the data volumes measured with incident energies $E_{\rm i} = 51$, 111 and 303\,meV. To extract the magnetic dispersion in a form suitable for fitting to a model we  performed a large number of such constant-energy cuts at wavevectors along the reciprocal-space paths indicated in Fig.~\ref{fig:spins}(b). The peaks in these plus some additional constant-wavevector cuts were fitted with Gaussian functions on a linear background. The peak centers determined this way are plotted along the high symmetry directions in Fig.~\ref{fig:calcdisp}.

An interesting behaviour is observed along the magnetic zone boundary: the energy of the magnon branch is not constant but varies by approximately 1.5\,meV. As discussed below, this is significant because a dispersion along the zone boundary indicates a need to go beyond a linear spin-wave model with nearest-neighbour interactions only. To emphasise this effect, we show in Fig.~\ref{fig:MAPS_MXM}(a) the energy and (b) the integrated intensity of the magnon peak along the entire length of a zone boundary (X$\Sigma$X). The maximum in the dispersion at $\Sigma$ is seen to coincide with a minimum in its intensity.  Because the dispersion surface forms a ridge along the zone boundary care was taken to select an appropriately-sized box in $\bf Q$ over which to average the data so as to avoid systematic errors from the curvature of the dispersion surface while at the same time having good enough statistics to extract the peak energies and integrated intensities. Figure~\ref{fig:MAPS_MXM}(c) shows energy cuts taken at an X-point and a $\Sigma$-point to illustrate the difference between the magnon peaks at the zone corner and zone edge.

\section{Analysis and discussion}

% Simulated dispersion
\begin{figure}
\includegraphics[width=\columnwidth] {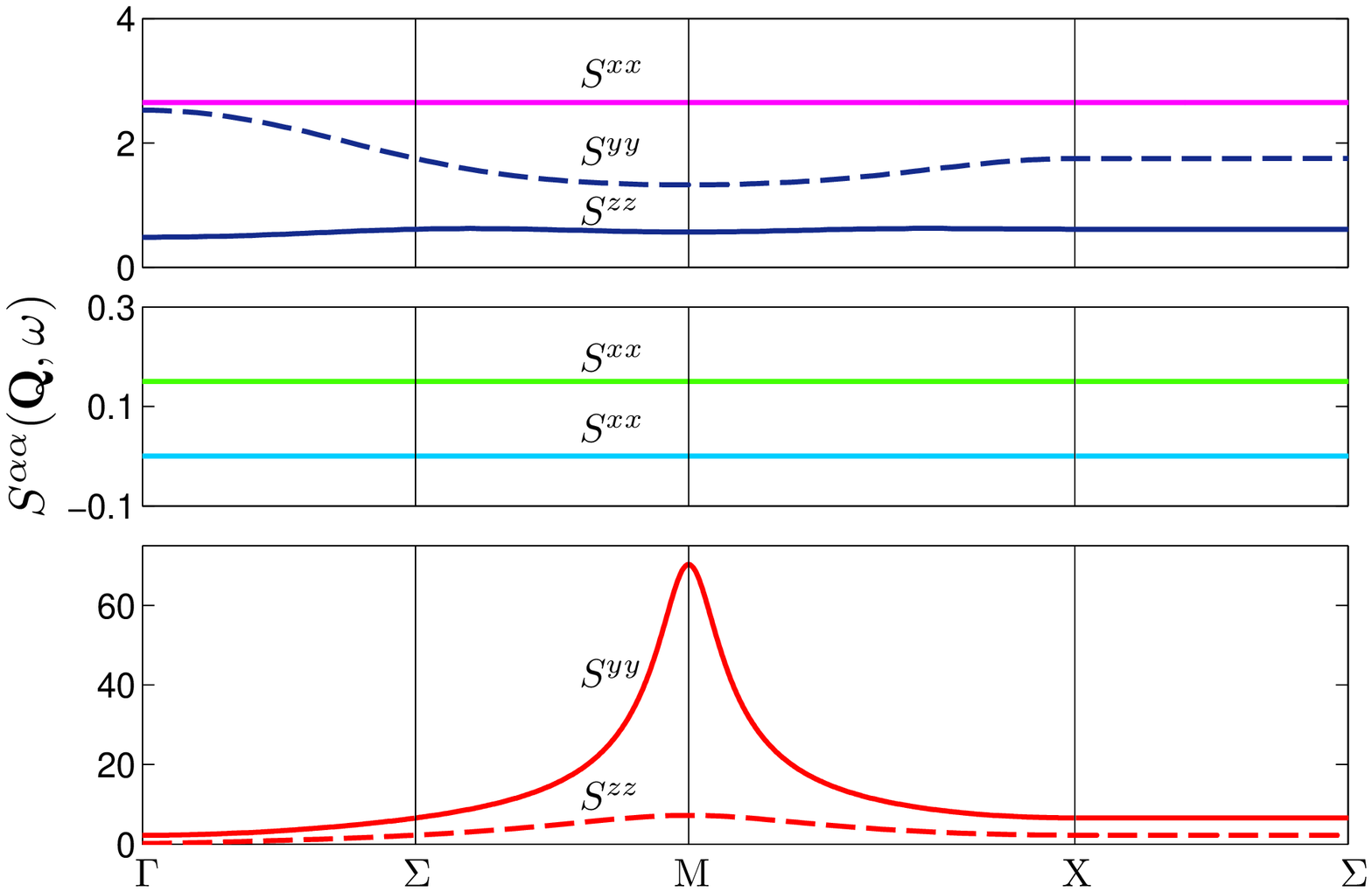}
\includegraphics[width=\columnwidth] {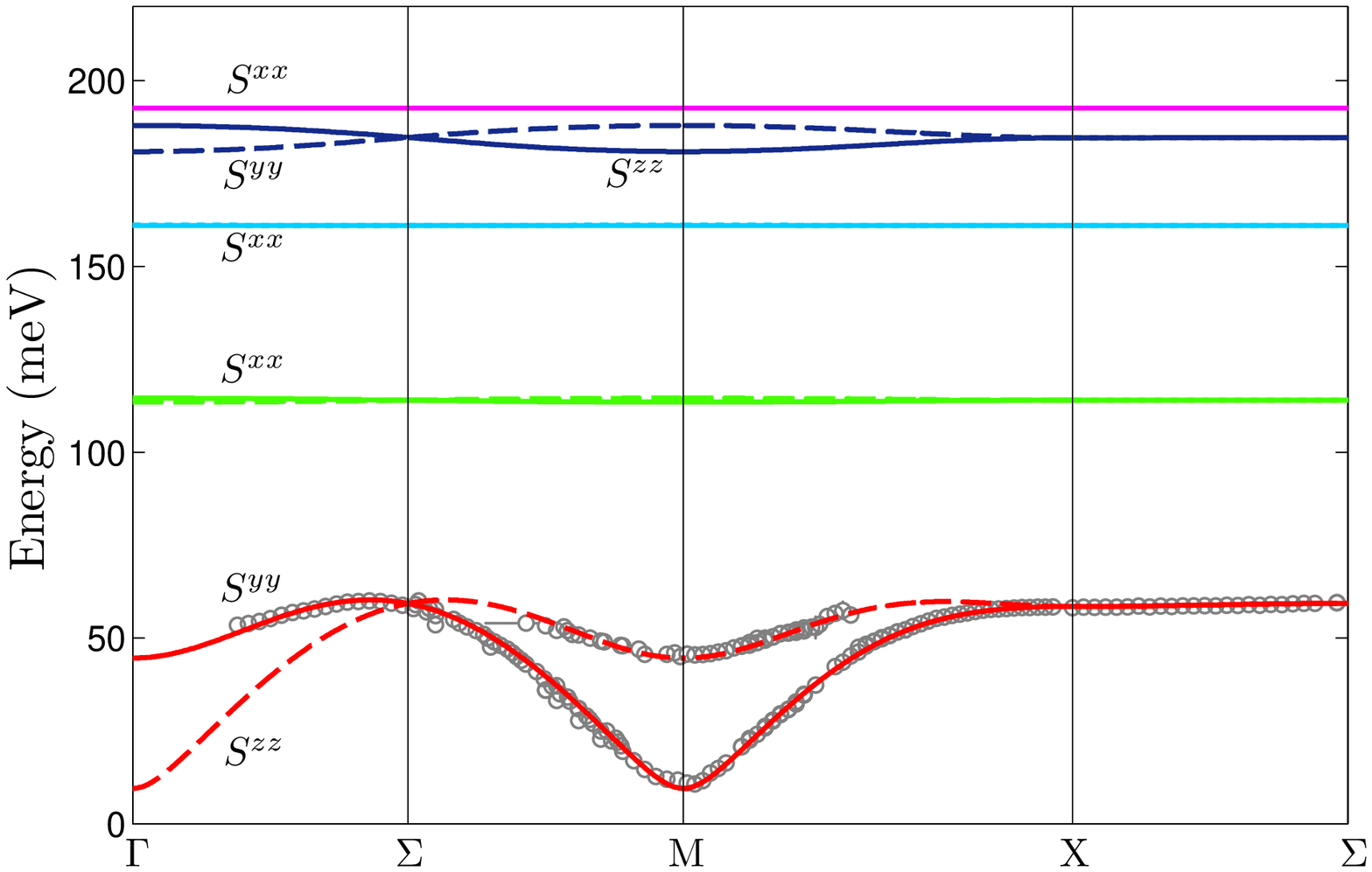}
\caption{  The lower figure shows the dispersion of the magnetic excitations of La$_2$CoO$_4$ along high symmetry directions in the 2D Brillouin zone defined in Fig.~\ref{fig:spins}. Open circles are points extracted from cuts through the measured data volume. The lines show the dispersion of the modes calculated with the many-level spin-wave model described in the text. The upper figure shows the response functions $S^{\alpha\alpha}$ for each mode calculated from the many-level model. The normalization of the response functions is per formula unit of La$_2$CoO$_4$.
\label{fig:calcdisp}}
\end{figure}

% Many-level model
Magnetism in Co$^{2+}$ compounds such as La$_2$CoO$_4$ is generally influenced to a significant degree by unquenched orbital angular momentum which is responsible for, among other things, the strong anisotropy in the susceptibility observed in many such compounds. In a recent study\cite{helme-prb-2009} of the magnetic excitations in the half-doped cobaltate La$_{1.5}$Sr$_{0.5}$CoO$_4$, which is also an antiferromagnet, we developed a model to describe the magnetic spectrum including both the spin and orbital angular momentum of the Co$^{2+}$ in the high-spin configuration ($3d^7$, $S=3/2$, $L = 3$). The model is an advance over conventional (spin-only) spin-wave theory in that it includes level-mixing within the $^{2S+1}L$ term caused by the ligand and exchange fields, and hence the parameters that describe the single-ion anisotropy and exchange interactions are physically realistic. As far as the magnetic spectrum is concerned, the admixture of basis states means that excitations to levels above the first excited single-ion level can propagate and can be observed by neutron scattering. Moreover, the orbital component of the single-ion states needs to be included for an accurate calculation of the neutron cross section.

The model employs the Hamiltonian
\begin{eqnarray}
{\mathcal H}& = & \sum_{\langle jk\rangle}J_{jk}{\bf S}_j \cdot {\bf
S}_k \nonumber \\ & & + \sum_j\left[\sum_{l,m} B_l^m O_l^m({\bf L}_j) + \lambda {\bf
L}_j
\cdot {\bf S}_j + {\bf H}^{\rm a}_j\cdot  {\bf S}_j\right]. \label{eq:Hmany}
\end{eqnarray}
The first term describes an isotropic Heisenberg exchange interaction between pairs of  $S=3/2$ spins. For La$_2$CoO$_4$ we include only the nearest-neighbour and next-nearest-neighbour exchange interactions $J$, $J_1$ and $J_2$, as defined in Fig.~\ref{fig:spins}. The remaining terms in (\ref{eq:Hmany}) are single-ion terms.  The first of these represents the crystal (ligand) field acting on the Co$^{2+}$ ions. The $O_l^m$ are Stevens operator-equivalents with $B_l^m$ the corresponding crystal-field parameters. The axially-distorted octahedral crystal field from the neighbouring O$^{2-}$ ions is described by the operators $O_2^0$, $O_4^0$ and $O_4^4$. We kept the same values for the parameters $B^0_4$ and $B^4_4$ as found for La$_{1.5}$Sr$_{0.5}$CoO$_4$ in Ref.~\onlinecite{helme-prb-2009}: $B^0_4 = -1.35$\,meV and  $B^4_4 = -8.00$\,meV. These are estimated from a point-charge calculation and scaled to match the cubic crystal field splitting observed in CoO.\cite{kant-prb-2008} The parameter $B^0_2$ controls the out-of-plane anisotropy and was adjusted to obtain a good fit to the magnetic spectrum. Its final value (see below) differs from that deduced for La$_{1.5}$Sr$_{0.5}$CoO$_4$ by only $\sim$10\%. The term $\lambda {\bf L} \cdot {\bf S}$ is the spin-orbit coupling. The coupling constant $\lambda=-18.7$\,meV used here has been deduced from reflectivity measurements of CoO by optical spectroscopy.\cite{kant-prb-2008} The final term ${\bf H}^{\rm a}\cdot {\bf S}$ represents a small uniaxial anisotropy which defines the in-plane orientation of the moments and produces a spin gap at the $\Gamma$-point (and, equivalently, the M-point). We chose the moments to lie along the $x$ axis, and to achieve this the anisotropy field ${\bf H}^{\rm a}$ points along $+x$ on one of the antiferromagnetic sublattices and along $-x$ on the other.

The partial differential scattering cross-section depends on the response functions $S^{\alpha\alpha}({\bf Q},\omega)$ describing $\alpha\alpha$ magnetic correlations. In the dipole approximation the relation is,\cite{squires-book}
%\begin{eqnarray}
\begin{equation}
\frac{k_{\rm i}}{k_{\rm f}}\frac{{\rm d}^2\sigma}{{\rm d}\Omega{\rm d}E_{\rm f}}
 =
\left(\frac{\gamma r_0}{2}
\right)^{\hspace{-2pt} 2} f^2(Q)\, {\rm e}^{-2W}%\nonumber\\
% & & \quad \times
\sum_{\alpha}(1-\hat{Q}_{\alpha}^2)S^{\alpha\alpha}({\bf
Q},\omega),\label{eq:cross-sec}
%\end{eqnarray}
\end{equation}
where
\begin{equation}
S^{\alpha\alpha}({\bf Q},\omega) =
\sum_j |\langle j|M^\alpha({\bf Q})|0\rangle|^2 \delta[\omega - \omega_j({\bf Q})].\label{eq:response}
\end{equation}
Here, $k_{\rm i}$ and $k_{\rm f}$ are initial and final neutron wavevectors, $(\gamma r_0/2)^2 = 72.8$\,mb, $f(Q)$ is the dipole magnetic form factor of Co$^{2+}$, ${\rm e}^{-2W}$ is the Debye-Waller factor which is close to unity at low temperatures, and $\hat{Q}_\alpha = Q_\alpha/|{\bf Q}|$ is the $\alpha$ component of a unit vector in the direction of $\bf Q$.  The response function (per La$_2$CoO$_4$ f.u.) described in Eq.~(\ref{eq:response}) takes into account both the spin and orbital magnetization ${\bf M} = -({\bf L} + 2{\bf S})$ in the transition matrix element connecting the ground state to an excited mode $j$. The procedure to diagonalise the Hamiltonian (\ref{eq:Hmany}) to obtain the dispersion and response functions of the magnetic modes is described in detail in Ref.\ \onlinecite{helme-prb-2009}.

The parameters of the model were refined from a fit to the measured dispersion carried out by a simulated-annealing algorithm. Because of the computer time required to diagonalise the Hamiltonian for the complete set of $2\times \{(2L+1)(2S+1)-1\} = 54$ excited states (twice the number of single-ion excited states because we have two magnetic sublattices) we restricted the number of observables included in the fit to just enough to represent all the important features of the data, including the high-energy signal at $\sim$190\,meV. The parameters varied in the fit were $B^0_2$, $J$, $J_1$, $J_2$, and $H^{\rm a}$. The best fit was achieved with parameters $B^0_2 = 14.6(1)$\,meV, $J = 9.69(2)$\,meV, $J_1 = 0.14(2)$\,meV, $J_2 = 0.43(1)$\,meV, and $H^{\rm a} = 0.66(6)$\,meV. The calculated dispersion and response functions of the magnetic modes are shown in Fig.~\ref{fig:calcdisp} together with the full set of data points for the lowest energy modes determined from the measurements. The agreement is seen to be very good. The fit indicates that the next-nearest-neighbour exchange constants $J_1$ and $J_2$ are very small but not zero. As a test, we repeated the fit with $J_1$ and $J_2$ fixed to zero and found that the quality of best fit worsened, as indicated by the standard goodness-of-fit parameter $\chi^2$ per degree of freedom which increased from 4.5 to 11.1. Therefore, we believe that the obtained values of $J_1$ and $J_2$, though small, are significant.

% MXM dispersion line
\begin{figure}
\includegraphics[width=\columnwidth]{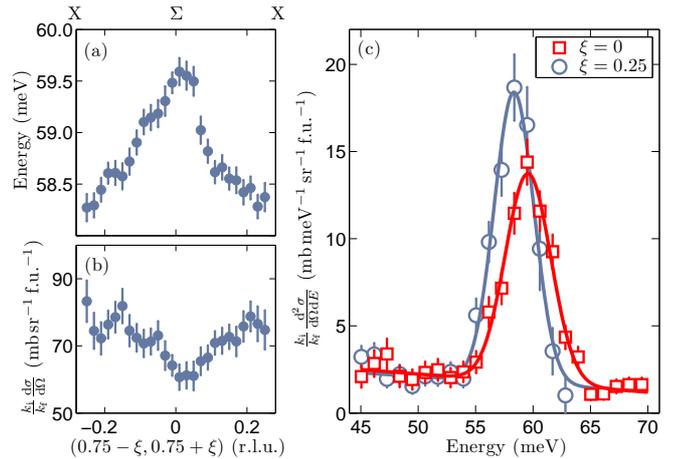}
\caption{  Dispersion of the magnon peak along the magnetic zone boundary in La$_2$CoO$_4$. Variation of (a) the peak position and (b) the integrated intensity of the magnon peak in constant-$\bf Q$ cuts. (c) Constant-$\bf Q$ cuts at ${\bf Q} = (0.75,0.75)$ and $(0.5,1)$ fitted with Gaussian lineshapes. The data are from the run with incident energy 111\,meV and sample temperature 6\,K.
\label{fig:MAPS_MXM}}
\end{figure}

To further visualise and assess the model we calculated intensity maps and cuts to simulate those obtained from the experiment. Figures~\ref{fig:MAPS_ExcitonQ} and \ref{fig:cuts} show the simulations alongside the corresponding experimental data. The quantity plotted is $(k_{\rm i}/k_{\rm f}) {\rm d}^2\sigma/{\rm d}\Omega{\rm d}E_{\rm f}$ per formula unit (f.u.), i.e., the partial differential cross section multiplied by a factor $k_{\rm i}/k_{\rm f}$ as defined in Eq.~(\ref{eq:cross-sec}). The dipole magnetic form factor of Co$^{2+}$ and the $\bf Q$ orientation factor that determines the weighting of the different response functions are included in the simulated spectra. The simulations also take into account a number of other experimental factors: (i) we averaged over a 50:50 mixture of equivalent magnetic domains in which the ordered moments point along the $x$ and $y$ axes, respectively; (ii) the spectra are broadened in energy and wavevector by the estimated resolution of the MAPS spectrometer (see Sec.~\ref{sec:expt}); (iii) we included an estimate of the absorption and self-shielding of the neutron beam by the sample, which reduces the intensity by a factor of typically 0.65--0.80 depending on the incident neutron energy and $\hbar \omega$. An additional scale factor of 0.4 was applied uniformly to all calculated spectra in order to match the measured absolute scattering intensity.

% Note on the fit results

The simulations show that the model provides a very good description of the entire observed spectrum of La$_2$CoO$_4$. The relative intensities of the magnetic excitations are reproduced to within 10--20\,\%, including the band of scattering at $\sim$190\,meV, which from Fig.~\ref{fig:calcdisp} is seen to originate from a mode with longitudinal ($xx$) character together with some less-intense transverse modes. Magnetic excitations are also present in the model at around $\sim$115 and $\sim$165\,meV but are predicted to carry negligible spectral weight and are not observed --- see Fig.~\ref{fig:cuts}(c). The additional scale factor of 0.4 needed to match the absolute intensity is similar to that required for La$_{1.5}$Sr$_{0.5}$CoO$_4$ (Ref.~\onlinecite{helme-prb-2009}). It is accounted for partly by the size of the ordered moment. The observed ordered moment is 2.9\,$\mu_{\rm B}$ (Ref.~\onlinecite{yamada-prb-1989}) whereas the ordered moment in the (ionic) model is 4.1\,$\mu_{\rm B}$. The difference between observed and calculated moments may be an effect of covalency, which would also modify the magnetic form factor relative to the free ion form factor in such a way that could cause an additional reduction in intensity, as recently found in a cuprate chain compound.\cite{walters-natphys-2009}

% Linear spin-wave model

For reference, we also compared the low-energy part of the spectrum ($\hbar\omega < 60$\,meV) with standard linear spin-wave theory for an effective spin--$\frac{1}{2}$ antiferromagnet, which neglects the orbital component of the modes. We used the same model as described in Ref.\ \onlinecite{helme-prb-2009} in which the magnetic anisotropy is described by anisotropic nearest-neighbour exchange interactions $J_x = J(1+\epsilon)$, $J_y = J$, and $J_z = J(1-\delta)$. The parameters $\epsilon$ and $\delta$ control the in-plane and out-of-plane anisotropy, respectively. The more distant interactions $J_1$ and $J_2$ were included too, but because they are relatively small we treated these as isotropic. We found that the lower energy modes can be well described by this model. In fact, an equally good description of the data (as reflected in the value of $\chi^2$) could be found with sets of parameters in which $J_1$ and $J_2$ are both positive or both negative: (i) $J = 9.89(1)$\,meV, $J_1 = 0.04(1)$\,meV, $J_2 = 0.13(1)$\,meV, $\epsilon = 0.013(1)$, $\delta = 0.283(4)$, or (ii) $J = 8.30(6)$\,meV, $J_1 = -0.35(2)$\,meV, $J_2 = -0.63(3)$\,meV, $\epsilon = 0.024(1)$, $\delta = 0.383(5)$. By contrast, the spin--orbital many-level model clearly favors the case with $J_1$ and $J_2$ both positive. The spin--orbital model can discriminate the two cases because of the inclusion of the higher excited levels. Only the parameter set with $J_1$ and $J_2$ both positive fits the low energy modes ($E<60$\,meV) {\it and} reproduces the peak in the spectrum at $\sim$190\,meV and absence of any other measurable peaks between 60 and 250\,meV. Another drawback of the effective spin--$\frac{1}{2}$ linear spin-wave model is that the intensities are not accurately described because of the neglect of the orbital degrees of freedom.

It is interesting to compare the magnetic spectrum of La$_2$CoO$_4$ with that of other two-dimensional, square-lattice, antiferromagnetic insulators, particularly in relation to the anomalous dispersion along the zone boundary. By ``anomalous", we mean that the zone-boundary dispersion cannot be described within the framework of an antiferromagnetic spin-wave model in the linear approximation with only nearest-neighbour interactions. Inclusion of (i) interactions with more distant neighbours, or (ii) terms beyond the linear approximation, are two ways in which a zone-boundary dispersion can be obtained. Other layered antiferromagnets which exhibit zone-boundary dispersion include La$_2$CuO$_4$ (Refs.\ \onlinecite{coldea-prl-2001,headings-arxiv}), Sr$_2$Cu$_3$O$_4$Cl$_2$ (Ref.\ \onlinecite{kim-prl-1999}) and Cu(DCOO)$_2\, \cdot$ 4D$_2$O (CFTD, Refs.\ \onlinecite{ronnow-prl-2001,christensen-pnas-2007}). These are all highly two-dimensional, $S=\frac{1}{2}$ Heisenberg antiferromagnets with almost isotropic interactions, and it is thought that the zone-boundary dispersion is caused by non-linear terms in the nearest-neighbour Heisenberg model. For example, in La$_2$CuO$_4$ a model with a four-spin ring exchange was employed,\cite{coldea-prl-2001} and for CFTD a resonating-valence-bond model describing entangled spin-dimer states was proposed to explain the data.\cite{christensen-pnas-2007} Interestingly, the behaviour along the zone boundary is different in these two materials: in La$_2$CuO$_4$ both the energy and intensity are higher at X than at $\Sigma$, whereas in CFTD both the energy and intensity are higher at $\Sigma$ than at X. In La$_2$CoO$_4$, on the other hand, the energy is a maximum at $\Sigma$ while the intensity is a maximum at X (see Fig.~\ref{fig:MAPS_MXM}).  By contrast, there is virtually no zone-boundary dispersion at all in $S=5/2$ square-lattice system Rb$_2$MnF$_4$, Ref.\ \onlinecite{huberman-prb-2005}. In our analysis of La$_2$CoO$_4$ we found that although the zone boundary dispersion can be satisfactorily reproduced with an appropriate choice of $J_1$ and $J_2$, the corresponding intensity does not have the deep minimum at $\Sigma$ found in the experiment [Fig.~\ref{fig:MAPS_MXM}(b)].  Therefore, whether the zone boundary dispersion of La$_2$CoO$_4$ is due to interactions with more distant spins or arises from quantum effects in a non-linear nearest-neighbour model remains an open question.

% Temperature dependence of the dispersion
\begin{figure}
\includegraphics[width=\columnwidth, bb = 0 221 605 502,clip] {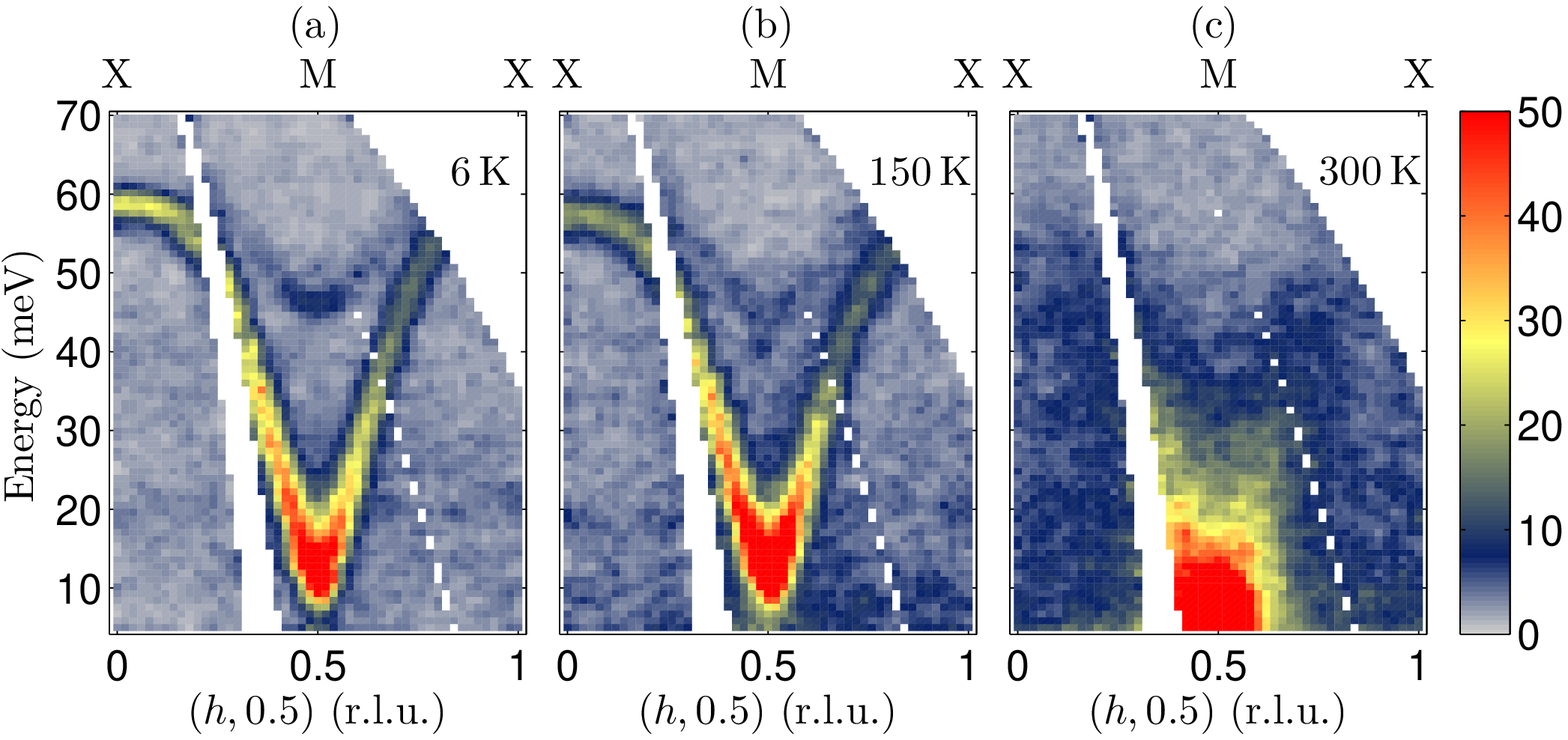}\vspace{0.5cm}
\includegraphics[width=\columnwidth]
{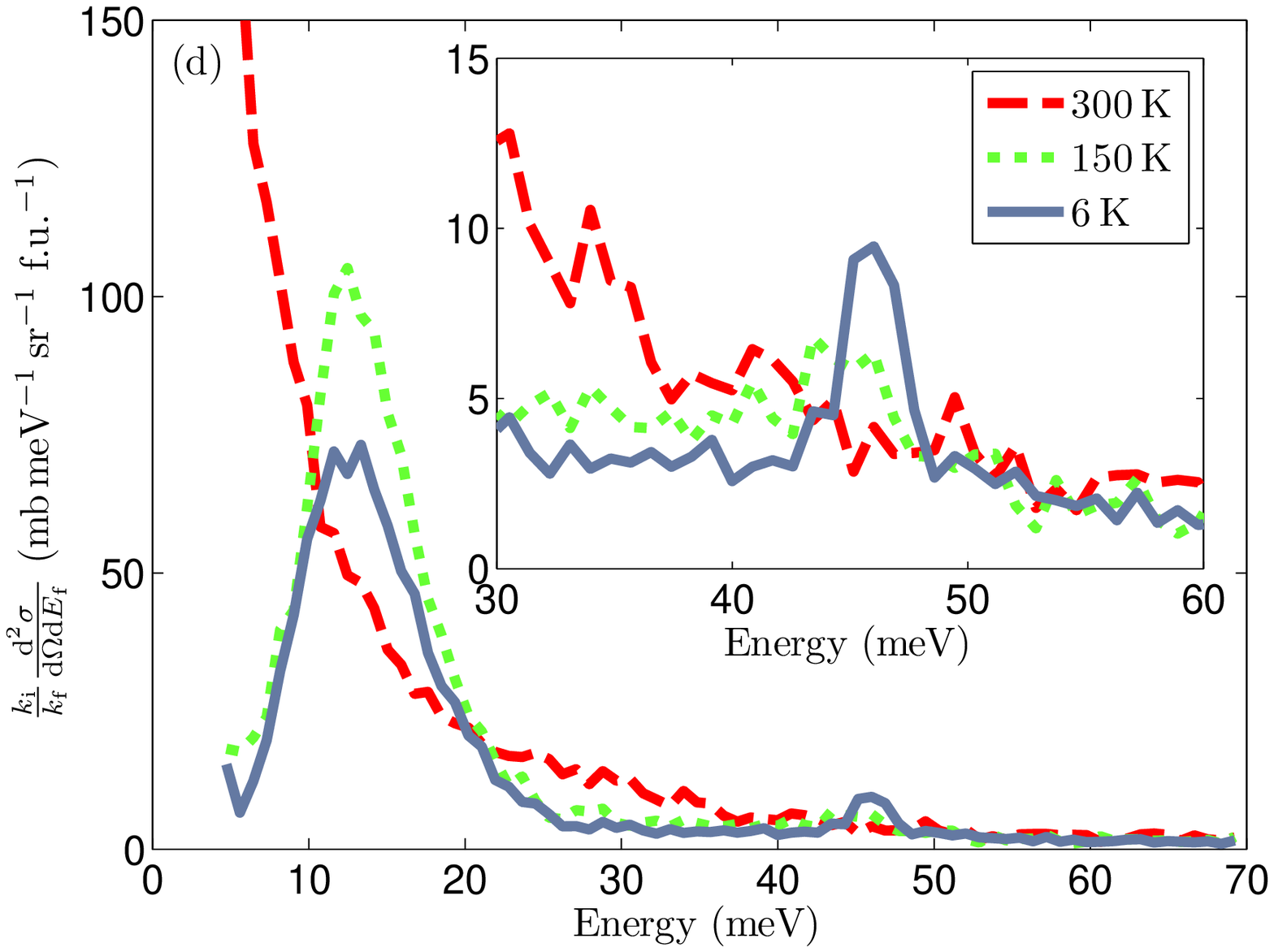}
\caption{  Temperature dependence of the magnetic spectrum of La$_2$CoO$_4$. The upper panels (a)--(c) show intensity maps measured at 6, 150 and 300\,K along the $(h,0.5)$ direction. (d) Magnetic spectrum at the antiferromagnetic ordering wavevector $(0.5,0.5)$ (M-point) measured at 6, 150 and 300\,K. Inset: temperature evolution of the higher-energy magnon mode.
\label{fig:AFM_T}}
\end{figure}

Finally, we consider the temperature dependence of the magnetic spectrum. Figures~\ref{fig:AFM_T}(a)--(c) show maps of the magnetic scattering measured at $T=6$, 150 and 300\,K, and Fig.~\ref{fig:AFM_T}(d) displays constant-{\bf Q} cuts at the magnetic zone center for the same temperatures. On increasing the temperature from 6 to 150\,K the 11\,meV peak increases in intensity due to the increasing thermal population but remains at the same energy, while the 46\,meV peak broadens and shifts to lower energy [Fig.~\ref{fig:AFM_T}(d) inset]. Although La$_2$CoO$_4$ undergoes a first-order phase transition coincident with a magnetic reorientation at $T_2 \approx 125$\,K, the in-plane lattice parameters in the LTO phase differ only slightly from those in the LTT phase, and the change in the magnetic structure only affects the stacking along the $c$ axis. It is not surprising, therefore, that the transition does not significantly affect the magnetic spectrum.  At $T=300$\,K, the spectrum has become quasielastic and there are no longer any sharp inelastic peaks. This indicates the absence of long-range magnetic correlations for $T>T_{\rm N}$.

\section{Conclusion}

We have measured the excitation spectrum of single-crystal La$_2$CoO$_4$, an excellent realization of a two-dimensional Heisenberg antiferromagnet. We have combined the experimental results with numerical simulations to achieve a very good description of the magnetic spectrum throughout the entire Brillouin zone, up to an energy of 250\,meV. The magnetic anisotropy is strongly XY-like, but a small uniaxial anisotropy is present which will make the low temperature magnetic properties Ising-like. An anomalous dispersion along the antiferromagnetic zone boundary is observed and can be reproduced by including exchange interactions beyond the nearest-neighbours but which could also be a manifestation of quantum fluctuations in a nearest-neighbour model.

\begin{acknowledgments}
We wish to acknowledge helpful discussions with B.~Keimer, B.~Roessli, G.~L.~Pascut and G.~Johnstone in the preparation of this work. We thank R.~A.~Ewings for valuable assistance with the data analysis, and T.~G.~ Perring for help with the experiments. P.B. is grateful for the provision of a studentship from the UK Engineering and Physical Sciences Research Council.
\end{acknowledgments}

% You should use BibTeX and apsrev.bst for references
% Choosing a journal automatically selects the correct APS
% BibTeX style file (bst file), so only uncomment the line
% below if necessary.
\bibliographystyle{apsrev4-1}

\bibliography{biblio}

\begin{thebibliography}{30}%
\makeatletter
\providecommand \@ifxundefined [1]{%
 \@ifx{#1\undefined}
}%
\providecommand \@ifnum [1]{%
 \ifnum #1\expandafter \@firstoftwo
 \else \expandafter \@secondoftwo
 \fi
}%
\providecommand \@ifx [1]{%
 \ifx #1\expandafter \@firstoftwo
 \else \expandafter \@secondoftwo
 \fi
}%
\providecommand \natexlab [1]{#1}%
\providecommand \enquote  [1]{``#1''}%
\providecommand \bibnamefont  [1]{#1}%
\providecommand \bibfnamefont [1]{#1}%
\providecommand \citenamefont [1]{#1}%
\providecommand \href@noop [0]{\@secondoftwo}%
\providecommand \href [0]{\begingroup \@sanitize@url \@href}%
\providecommand \@href[1]{\@@startlink{#1}\@@href}%
\providecommand \@@href[1]{\endgroup#1\@@endlink}%
\providecommand \@sanitize@url [0]{\catcode `\\12\catcode `\$12\catcode
  `\&12\catcode `\#12\catcode `\^12\catcode `\_12\catcode `\%12\relax}%
\providecommand \@@startlink[1]{}%
\providecommand \@@endlink[0]{}%
\providecommand \url  [0]{\begingroup\@sanitize@url \@url }%
\providecommand \@url [1]{\endgroup\@href {#1}{\urlprefix }}%
\providecommand \urlprefix  [0]{URL }%
\providecommand \Eprint [0]{\href }%
\providecommand \doibase [0]{http://dx.doi.org/}%
\providecommand \selectlanguage [0]{\@gobble}%
\providecommand \bibinfo  [0]{\@secondoftwo}%
\providecommand \bibfield  [0]{\@secondoftwo}%
\providecommand \translation [1]{[#1]}%
\providecommand \BibitemOpen [0]{}%
\providecommand \bibitemStop [0]{}%
\providecommand \bibitemNoStop [0]{.\EOS\space}%
\providecommand \EOS [0]{\spacefactor3000\relax}%
\providecommand \BibitemShut  [1]{\csname bibitem#1\endcsname}%
\let\auto@bib@innerbib\@empty
%</preamble>
\bibitem [{\citenamefont {Bednorz}\ and\ \citenamefont
  {M\"{u}ller}(1986)}]{bednorz-1986}%
  \BibitemOpen
  \bibfield  {author} {\bibinfo {author} {\bibfnamefont {J.~G.}\ \bibnamefont
  {Bednorz}}\ and\ \bibinfo {author} {\bibfnamefont {K.~A.}\ \bibnamefont
  {M\"{u}ller}},\ }\href@noop {} {\bibfield  {journal} {\bibinfo  {journal} {Z.
  Phys. B}\ }\textbf {\bibinfo {volume} {64}},\ \bibinfo {pages} {189}
  (\bibinfo {year} {1986})}\BibitemShut {NoStop}%
\bibitem [{\citenamefont {Tranquada}\ \emph {et~al.}(1995)\citenamefont
  {Tranquada}, \citenamefont {Sternleib}, \citenamefont {Axe}, \citenamefont
  {Nakamura},\ and\ \citenamefont {Uchida}}]{tranquada-nature-1995}%
  \BibitemOpen
  \bibfield  {author} {\bibinfo {author} {\bibfnamefont {J.~M.}\ \bibnamefont
  {Tranquada}}, \bibinfo {author} {\bibfnamefont {B.~J.}\ \bibnamefont
  {Sternleib}}, \bibinfo {author} {\bibfnamefont {J.~D.}\ \bibnamefont {Axe}},
  \bibinfo {author} {\bibfnamefont {Y.}~\bibnamefont {Nakamura}}, \ and\
  \bibinfo {author} {\bibfnamefont {S.}~\bibnamefont {Uchida}},\ }\href@noop {}
  {\bibfield  {journal} {\bibinfo  {journal} {Nature (London)}\ }\textbf
  {\bibinfo {volume} {375}},\ \bibinfo {pages} {561} (\bibinfo {year}
  {1995})}\BibitemShut {NoStop}%
\bibitem [{\citenamefont {Chen}\ \emph {et~al.}(1993)\citenamefont {Chen},
  \citenamefont {Cheong},\ and\ \citenamefont {Cooper}}]{chen-prl-1993}%
  \BibitemOpen
  \bibfield  {author} {\bibinfo {author} {\bibfnamefont {C.~H.}\ \bibnamefont
  {Chen}}, \bibinfo {author} {\bibfnamefont {S.~W.}\ \bibnamefont {Cheong}}, \
  and\ \bibinfo {author} {\bibfnamefont {A.~S.}\ \bibnamefont {Cooper}},\
  }\href@noop {} {\bibfield  {journal} {\bibinfo  {journal} {Phys. Rev. Lett.}\
  }\textbf {\bibinfo {volume} {71}},\ \bibinfo {pages} {2461} (\bibinfo {year}
  {1993})}\BibitemShut {NoStop}%
\bibitem [{\citenamefont {Tranquada}\ \emph {et~al.}(1994)\citenamefont
  {Tranquada}, \citenamefont {Buttrey}, \citenamefont {Sachan},\ and\
  \citenamefont {Lorenzo}}]{tranquada-prl-1994}%
  \BibitemOpen
  \bibfield  {author} {\bibinfo {author} {\bibfnamefont {J.~M.}\ \bibnamefont
  {Tranquada}}, \bibinfo {author} {\bibfnamefont {D.~J.}\ \bibnamefont
  {Buttrey}}, \bibinfo {author} {\bibfnamefont {V.}~\bibnamefont {Sachan}}, \
  and\ \bibinfo {author} {\bibfnamefont {J.~E.}\ \bibnamefont {Lorenzo}},\
  }\href@noop {} {\bibfield  {journal} {\bibinfo  {journal} {Phys. Rev. Lett.}\
  }\textbf {\bibinfo {volume} {73}},\ \bibinfo {pages} {1003} (\bibinfo {year}
  {1994})}\BibitemShut {NoStop}%
\bibitem [{\citenamefont {Wochner}\ \emph {et~al.}(1998)\citenamefont
  {Wochner}, \citenamefont {Tranquada}, \citenamefont {Buttrey},\ and\
  \citenamefont {Sachan}}]{wochner-prb-1998}%
  \BibitemOpen
  \bibfield  {author} {\bibinfo {author} {\bibfnamefont {P.}~\bibnamefont
  {Wochner}}, \bibinfo {author} {\bibfnamefont {J.~M.}\ \bibnamefont
  {Tranquada}}, \bibinfo {author} {\bibfnamefont {D.~J.}\ \bibnamefont
  {Buttrey}}, \ and\ \bibinfo {author} {\bibfnamefont {V.}~\bibnamefont
  {Sachan}},\ }\href@noop {} {\bibfield  {journal} {\bibinfo  {journal} {Phys.
  Rev. B}\ }\textbf {\bibinfo {volume} {57}},\ \bibinfo {pages} {1066}
  (\bibinfo {year} {1998})}\BibitemShut {NoStop}%
\bibitem [{\citenamefont {Yoshizawa}\ \emph {et~al.}(2000)\citenamefont
  {Yoshizawa}, \citenamefont {Kakeshita}, \citenamefont {Kajimoto},
  \citenamefont {Tanabe}, \citenamefont {Katsufuji},\ and\ \citenamefont
  {Tokura}}]{yoshizawa-prb-2000}%
  \BibitemOpen
  \bibfield  {author} {\bibinfo {author} {\bibfnamefont {H.}~\bibnamefont
  {Yoshizawa}}, \bibinfo {author} {\bibfnamefont {T.}~\bibnamefont
  {Kakeshita}}, \bibinfo {author} {\bibfnamefont {R.}~\bibnamefont {Kajimoto}},
  \bibinfo {author} {\bibfnamefont {T.}~\bibnamefont {Tanabe}}, \bibinfo
  {author} {\bibfnamefont {T.}~\bibnamefont {Katsufuji}}, \ and\ \bibinfo
  {author} {\bibfnamefont {Y.}~\bibnamefont {Tokura}},\ }\href@noop {}
  {\bibfield  {journal} {\bibinfo  {journal} {Phys. Rev. B}\ }\textbf {\bibinfo
  {volume} {61}},\ \bibinfo {pages} {R854} (\bibinfo {year}
  {2000})}\BibitemShut {NoStop}%
\bibitem [{\citenamefont {Boothroyd}\ \emph
  {et~al.}(2003{\natexlab{a}})\citenamefont {Boothroyd}, \citenamefont
  {Freeman}, \citenamefont {Prabhakaran}, \citenamefont {Hiess}, \citenamefont
  {Enderle}, \citenamefont {Kulda},\ and\ \citenamefont
  {Altorfer}}]{boothroyd-prl-2003}%
  \BibitemOpen
  \bibfield  {author} {\bibinfo {author} {\bibfnamefont {A.~T.}\ \bibnamefont
  {Boothroyd}}, \bibinfo {author} {\bibfnamefont {P.~G.}\ \bibnamefont
  {Freeman}}, \bibinfo {author} {\bibfnamefont {D.}~\bibnamefont
  {Prabhakaran}}, \bibinfo {author} {\bibfnamefont {A.}~\bibnamefont {Hiess}},
  \bibinfo {author} {\bibfnamefont {M.}~\bibnamefont {Enderle}}, \bibinfo
  {author} {\bibfnamefont {J.}~\bibnamefont {Kulda}}, \ and\ \bibinfo {author}
  {\bibfnamefont {F.}~\bibnamefont {Altorfer}},\ }\href@noop {} {\bibfield
  {journal} {\bibinfo  {journal} {Phys. Rev. Lett.}\ }\textbf {\bibinfo
  {volume} {91}},\ \bibinfo {pages} {257201} (\bibinfo {year}
  {2003}{\natexlab{a}})}\BibitemShut {NoStop}%
\bibitem [{\citenamefont {Boothroyd}\ \emph
  {et~al.}(2003{\natexlab{b}})\citenamefont {Boothroyd}, \citenamefont
  {Prabhakaran}, \citenamefont {Freeman}, \citenamefont {Lister}, \citenamefont
  {Enderle}, \citenamefont {Hiess},\ and\ \citenamefont
  {Kulda}}]{boothroyd-prb-2003}%
  \BibitemOpen
  \bibfield  {author} {\bibinfo {author} {\bibfnamefont {A.~T.}\ \bibnamefont
  {Boothroyd}}, \bibinfo {author} {\bibfnamefont {D.}~\bibnamefont
  {Prabhakaran}}, \bibinfo {author} {\bibfnamefont {P.~G.}\ \bibnamefont
  {Freeman}}, \bibinfo {author} {\bibfnamefont {S.~J.~S.}\ \bibnamefont
  {Lister}}, \bibinfo {author} {\bibfnamefont {M.}~\bibnamefont {Enderle}},
  \bibinfo {author} {\bibfnamefont {A.}~\bibnamefont {Hiess}}, \ and\ \bibinfo
  {author} {\bibfnamefont {J.}~\bibnamefont {Kulda}},\ }\href@noop {}
  {\bibfield  {journal} {\bibinfo  {journal} {Phys. Rev. B}\ }\textbf {\bibinfo
  {volume} {67}},\ \bibinfo {pages} {100407(R)} (\bibinfo {year}
  {2003}{\natexlab{b}})}\BibitemShut {NoStop}%
\bibitem [{\citenamefont {Cwik}\ \emph {et~al.}(2009)\citenamefont {Cwik},
  \citenamefont {Benomar}, \citenamefont {Finger}, \citenamefont {Sidis},
  \citenamefont {Senff}, \citenamefont {Reuther}, \citenamefont {Lorenz},\ and\
  \citenamefont {Braden}}]{cwik-prl-2009}%
  \BibitemOpen
  \bibfield  {author} {\bibinfo {author} {\bibfnamefont {M.}~\bibnamefont
  {Cwik}}, \bibinfo {author} {\bibfnamefont {M.}~\bibnamefont {Benomar}},
  \bibinfo {author} {\bibfnamefont {T.}~\bibnamefont {Finger}}, \bibinfo
  {author} {\bibfnamefont {Y.}~\bibnamefont {Sidis}}, \bibinfo {author}
  {\bibfnamefont {D.}~\bibnamefont {Senff}}, \bibinfo {author} {\bibfnamefont
  {M.}~\bibnamefont {Reuther}}, \bibinfo {author} {\bibfnamefont
  {T.}~\bibnamefont {Lorenz}}, \ and\ \bibinfo {author} {\bibfnamefont
  {M.}~\bibnamefont {Braden}},\ }\href@noop {} {\bibfield  {journal} {\bibinfo
  {journal} {Phys. Rev. Lett.}\ }\textbf {\bibinfo {volume} {102}},\ \bibinfo
  {pages} {057201} (\bibinfo {year} {2009})}\BibitemShut {NoStop}%
\bibitem [{\citenamefont {Zaliznyak}\ \emph {et~al.}(2000)\citenamefont
  {Zaliznyak}, \citenamefont {Hill}, \citenamefont {Tranquada}, \citenamefont
  {Erwin},\ and\ \citenamefont {Moritomo}}]{zaliznyak-prl-2000}%
  \BibitemOpen
  \bibfield  {author} {\bibinfo {author} {\bibfnamefont {I.~A.}\ \bibnamefont
  {Zaliznyak}}, \bibinfo {author} {\bibfnamefont {J.~P.}\ \bibnamefont {Hill}},
  \bibinfo {author} {\bibfnamefont {J.~M.}\ \bibnamefont {Tranquada}}, \bibinfo
  {author} {\bibfnamefont {R.}~\bibnamefont {Erwin}}, \ and\ \bibinfo {author}
  {\bibfnamefont {Y.}~\bibnamefont {Moritomo}},\ }\href@noop {} {\bibfield
  {journal} {\bibinfo  {journal} {Phys. Rev. Lett.}\ }\textbf {\bibinfo
  {volume} {85}},\ \bibinfo {pages} {4353} (\bibinfo {year}
  {2000})}\BibitemShut {NoStop}%
\bibitem [{\citenamefont {Savici}\ \emph {et~al.}(2007)\citenamefont {Savici},
  \citenamefont {Zaliznyak}, \citenamefont {Gu},\ and\ \citenamefont
  {Erwin}}]{savici-prb-2007}%
  \BibitemOpen
  \bibfield  {author} {\bibinfo {author} {\bibfnamefont {A.~T.}\ \bibnamefont
  {Savici}}, \bibinfo {author} {\bibfnamefont {I.~A.}\ \bibnamefont
  {Zaliznyak}}, \bibinfo {author} {\bibfnamefont {G.~D.}\ \bibnamefont {Gu}}, \
  and\ \bibinfo {author} {\bibfnamefont {R.}~\bibnamefont {Erwin}},\
  }\href@noop {} {\bibfield  {journal} {\bibinfo  {journal} {Phys. Rev. B}\
  }\textbf {\bibinfo {volume} {75}},\ \bibinfo {pages} {184443} (\bibinfo
  {year} {2007})}\BibitemShut {NoStop}%
\bibitem [{\citenamefont {Horigane}\ \emph {et~al.}(2007)\citenamefont
  {Horigane}, \citenamefont {Hiraka}, \citenamefont {Uchida}, \citenamefont
  {Yamada},\ and\ \citenamefont {Akimitsu}}]{horigane-jpsj-2007}%
  \BibitemOpen
  \bibfield  {author} {\bibinfo {author} {\bibfnamefont {K.}~\bibnamefont
  {Horigane}}, \bibinfo {author} {\bibfnamefont {H.}~\bibnamefont {Hiraka}},
  \bibinfo {author} {\bibfnamefont {T.}~\bibnamefont {Uchida}}, \bibinfo
  {author} {\bibfnamefont {K.}~\bibnamefont {Yamada}}, \ and\ \bibinfo {author}
  {\bibfnamefont {J.}~\bibnamefont {Akimitsu}},\ }\href@noop {} {\bibfield
  {journal} {\bibinfo  {journal} {J. Phys. Soc. Jpn.}\ }\textbf {\bibinfo
  {volume} {76}},\ \bibinfo {pages} {114715} (\bibinfo {year}
  {2007})}\BibitemShut {NoStop}%
\bibitem [{\citenamefont {Hollmann}\ \emph {et~al.}(2008)\citenamefont
  {Hollmann}, \citenamefont {Haverkort}, \citenamefont {Cwik}, \citenamefont
  {Benomar}, \citenamefont {Reuther}, \citenamefont {Tanaka},\ and\
  \citenamefont {Lorenz}}]{hollmann-journphys-2008}%
  \BibitemOpen
  \bibfield  {author} {\bibinfo {author} {\bibfnamefont {N.}~\bibnamefont
  {Hollmann}}, \bibinfo {author} {\bibfnamefont {M.~W.}\ \bibnamefont
  {Haverkort}}, \bibinfo {author} {\bibfnamefont {M.}~\bibnamefont {Cwik}},
  \bibinfo {author} {\bibfnamefont {M.}~\bibnamefont {Benomar}}, \bibinfo
  {author} {\bibfnamefont {M.}~\bibnamefont {Reuther}}, \bibinfo {author}
  {\bibfnamefont {A.}~\bibnamefont {Tanaka}}, \ and\ \bibinfo {author}
  {\bibfnamefont {T.}~\bibnamefont {Lorenz}},\ }\href@noop {} {\bibfield
  {journal} {\bibinfo  {journal} {New J. Phys.}\ }\textbf {\bibinfo {volume}
  {10}},\ \bibinfo {pages} {023018} (\bibinfo {year} {2008})}\BibitemShut
  {NoStop}%
\bibitem [{\citenamefont {Chang}\ \emph {et~al.}(2009)\citenamefont {Chang},
  \citenamefont {Hu}, \citenamefont {Wu}, \citenamefont {Burnus}, \citenamefont
  {Hollmann}, \citenamefont {Benomar}, \citenamefont {Lorenz}, \citenamefont
  {Tanaka}, \citenamefont {Lin}, \citenamefont {Hsieh}, \citenamefont {Chen},\
  and\ \citenamefont {Tjeng}}]{chang-prl-2009}%
  \BibitemOpen
  \bibfield  {author} {\bibinfo {author} {\bibfnamefont {C.~F.}\ \bibnamefont
  {Chang}}, \bibinfo {author} {\bibfnamefont {Z.}~\bibnamefont {Hu}}, \bibinfo
  {author} {\bibfnamefont {H.}~\bibnamefont {Wu}}, \bibinfo {author}
  {\bibfnamefont {T.}~\bibnamefont {Burnus}}, \bibinfo {author} {\bibfnamefont
  {N.}~\bibnamefont {Hollmann}}, \bibinfo {author} {\bibfnamefont
  {M.}~\bibnamefont {Benomar}}, \bibinfo {author} {\bibfnamefont
  {T.}~\bibnamefont {Lorenz}}, \bibinfo {author} {\bibfnamefont
  {A.}~\bibnamefont {Tanaka}}, \bibinfo {author} {\bibfnamefont {H.~J.}\
  \bibnamefont {Lin}}, \bibinfo {author} {\bibfnamefont {H.~H.}\ \bibnamefont
  {Hsieh}}, \bibinfo {author} {\bibfnamefont {C.~T.}\ \bibnamefont {Chen}}, \
  and\ \bibinfo {author} {\bibfnamefont {L.~H.}\ \bibnamefont {Tjeng}},\
  }\href@noop {} {\bibfield  {journal} {\bibinfo  {journal} {Phys. Rev. Lett.}\
  }\textbf {\bibinfo {volume} {102}},\ \bibinfo {pages} {116401} (\bibinfo
  {year} {2009})}\BibitemShut {NoStop}%
\bibitem [{\citenamefont {Helme}\ \emph {et~al.}(2009)\citenamefont {Helme},
  \citenamefont {Boothroyd}, \citenamefont {Coldea}, \citenamefont
  {Prabhakaran}, \citenamefont {Frost}, \citenamefont {Keen}, \citenamefont
  {Regnault}, \citenamefont {Freeman}, \citenamefont {Enderle},\ and\
  \citenamefont {Kulda}}]{helme-prb-2009}%
  \BibitemOpen
  \bibfield  {author} {\bibinfo {author} {\bibfnamefont {L.~M.}\ \bibnamefont
  {Helme}}, \bibinfo {author} {\bibfnamefont {A.~T.}\ \bibnamefont
  {Boothroyd}}, \bibinfo {author} {\bibfnamefont {R.}~\bibnamefont {Coldea}},
  \bibinfo {author} {\bibfnamefont {D.}~\bibnamefont {Prabhakaran}}, \bibinfo
  {author} {\bibfnamefont {C.~D.}\ \bibnamefont {Frost}}, \bibinfo {author}
  {\bibfnamefont {D.~A.}\ \bibnamefont {Keen}}, \bibinfo {author}
  {\bibfnamefont {L.~P.}\ \bibnamefont {Regnault}}, \bibinfo {author}
  {\bibfnamefont {P.~G.}\ \bibnamefont {Freeman}}, \bibinfo {author}
  {\bibfnamefont {M.}~\bibnamefont {Enderle}}, \ and\ \bibinfo {author}
  {\bibfnamefont {J.}~\bibnamefont {Kulda}},\ }\href@noop {} {\bibfield
  {journal} {\bibinfo  {journal} {Phys. Rev. B}\ }\textbf {\bibinfo {volume}
  {80}},\ \bibinfo {pages} {134414} (\bibinfo {year} {2009})}\BibitemShut
  {NoStop}%
\bibitem [{\citenamefont {Yamada}\ \emph {et~al.}(1989)\citenamefont {Yamada},
  \citenamefont {Matsuda}, \citenamefont {Endoh}, \citenamefont {Keimer},
  \citenamefont {Birgeneau}, \citenamefont {Onodera}, \citenamefont {Mizusaki},
  \citenamefont {Matsuura},\ and\ \citenamefont {Shirane}}]{yamada-prb-1989}%
  \BibitemOpen
  \bibfield  {author} {\bibinfo {author} {\bibfnamefont {K.}~\bibnamefont
  {Yamada}}, \bibinfo {author} {\bibfnamefont {M.}~\bibnamefont {Matsuda}},
  \bibinfo {author} {\bibfnamefont {Y.}~\bibnamefont {Endoh}}, \bibinfo
  {author} {\bibfnamefont {B.}~\bibnamefont {Keimer}}, \bibinfo {author}
  {\bibfnamefont {R.~J.}\ \bibnamefont {Birgeneau}}, \bibinfo {author}
  {\bibfnamefont {S.}~\bibnamefont {Onodera}}, \bibinfo {author} {\bibfnamefont
  {J.}~\bibnamefont {Mizusaki}}, \bibinfo {author} {\bibfnamefont
  {T.}~\bibnamefont {Matsuura}}, \ and\ \bibinfo {author} {\bibfnamefont
  {G.}~\bibnamefont {Shirane}},\ }\href@noop {} {\bibfield  {journal} {\bibinfo
   {journal} {Phys. Rev. B}\ }\textbf {\bibinfo {volume} {39}},\ \bibinfo
  {pages} {2336} (\bibinfo {year} {1989})}\BibitemShut {NoStop}%
\bibitem [{\citenamefont {Perring}\ and\ \citenamefont
  {Frost}(2004)}]{perring-maps-2004}%
  \BibitemOpen
  \bibfield  {author} {\bibinfo {author} {\bibfnamefont {T.~G.}\ \bibnamefont
  {Perring}}\ and\ \bibinfo {author} {\bibfnamefont {C.~D.}\ \bibnamefont
  {Frost}},\ }\href@noop {} {\bibfield  {journal} {\bibinfo  {journal} {Neutron
  News}\ }\textbf {\bibinfo {volume} {15}},\ \bibinfo {pages} {30} (\bibinfo
  {year} {2004})}\BibitemShut {NoStop}%
\bibitem [{\citenamefont {Coldea}(2004)}]{coldea-mslice}%
  \BibitemOpen
  \bibfield  {author} {\bibinfo {author} {\bibfnamefont {R.}~\bibnamefont
  {Coldea}},\ }\href@noop {} {\bibfield  {journal} {\bibinfo  {journal}
  {MSLICE: a data analysis program for time-of-flight neutron spectrometers}\ }
  (\bibinfo {year} {2004})}\BibitemShut {NoStop}%
\bibitem [{\citenamefont {Lewandowski}\ \emph {et~al.}(1986)\citenamefont
  {Lewandowski}, \citenamefont {Beyerlein}, \citenamefont {Longo},\ and\
  \citenamefont {McCauley}}]{lewandowski-JAmCeramSoc-1986}%
  \BibitemOpen
  \bibfield  {author} {\bibinfo {author} {\bibfnamefont {J.~T.}\ \bibnamefont
  {Lewandowski}}, \bibinfo {author} {\bibfnamefont {R.~A.}\ \bibnamefont
  {Beyerlein}}, \bibinfo {author} {\bibfnamefont {J.~M.}\ \bibnamefont
  {Longo}}, \ and\ \bibinfo {author} {\bibfnamefont {R.~A.}\ \bibnamefont
  {McCauley}},\ }\href@noop {} {\bibfield  {journal} {\bibinfo  {journal} {J.
  Am. Ceram. Soc.}\ }\textbf {\bibinfo {volume} {69}},\ \bibinfo {pages} {699}
  (\bibinfo {year} {1986})}\BibitemShut {NoStop}%
\bibitem [{\citenamefont {Mohan~Ram}\ \emph {et~al.}(1988)\citenamefont
  {Mohan~Ram}, \citenamefont {Ganguly}, \citenamefont {Rao},\ and\
  \citenamefont {Honig}}]{mohan-MRB-1988}%
  \BibitemOpen
  \bibfield  {author} {\bibinfo {author} {\bibfnamefont {R.~A.}\ \bibnamefont
  {Mohan~Ram}}, \bibinfo {author} {\bibfnamefont {P.}~\bibnamefont {Ganguly}},
  \bibinfo {author} {\bibfnamefont {C.~N.~R.}\ \bibnamefont {Rao}}, \ and\
  \bibinfo {author} {\bibfnamefont {J.~M.}\ \bibnamefont {Honig}},\ }\href@noop
  {} {\bibfield  {journal} {\bibinfo  {journal} {Mat. Res. Bull.}\ }\textbf
  {\bibinfo {volume} {23}},\ \bibinfo {pages} {501} (\bibinfo {year}
  {1988})}\BibitemShut {NoStop}%
\bibitem [{\citenamefont {Kajitani}\ \emph {et~al.}(1990)\citenamefont
  {Kajitani}, \citenamefont {Hosoya}, \citenamefont {Hiraga},\ and\
  \citenamefont {Fukuda}}]{kajitani-jpsj-1990}%
  \BibitemOpen
  \bibfield  {author} {\bibinfo {author} {\bibfnamefont {T.}~\bibnamefont
  {Kajitani}}, \bibinfo {author} {\bibfnamefont {S.}~\bibnamefont {Hosoya}},
  \bibinfo {author} {\bibfnamefont {K.}~\bibnamefont {Hiraga}}, \ and\ \bibinfo
  {author} {\bibfnamefont {T.}~\bibnamefont {Fukuda}},\ }\href@noop {}
  {\bibfield  {journal} {\bibinfo  {journal} {J. Phys. Soc. Jpn.}\ }\textbf
  {\bibinfo {volume} {59}},\ \bibinfo {pages} {562} (\bibinfo {year}
  {1990})}\BibitemShut {NoStop}%
\bibitem [{\citenamefont {Kant}\ \emph {et~al.}(2008)\citenamefont {Kant},
  \citenamefont {Rudolf}, \citenamefont {Schrettle}, \citenamefont {Mayr},
  \citenamefont {Deisenhofer}, \citenamefont {Lunkenheimer}, \citenamefont
  {Eremin},\ and\ \citenamefont {Loidl}}]{kant-prb-2008}%
  \BibitemOpen
  \bibfield  {author} {\bibinfo {author} {\bibfnamefont {C.}~\bibnamefont
  {Kant}}, \bibinfo {author} {\bibfnamefont {T.}~\bibnamefont {Rudolf}},
  \bibinfo {author} {\bibfnamefont {F.}~\bibnamefont {Schrettle}}, \bibinfo
  {author} {\bibfnamefont {F.}~\bibnamefont {Mayr}}, \bibinfo {author}
  {\bibfnamefont {J.}~\bibnamefont {Deisenhofer}}, \bibinfo {author}
  {\bibfnamefont {P.}~\bibnamefont {Lunkenheimer}}, \bibinfo {author}
  {\bibfnamefont {M.~V.}\ \bibnamefont {Eremin}}, \ and\ \bibinfo {author}
  {\bibfnamefont {A.}~\bibnamefont {Loidl}},\ }\href@noop {} {\bibfield
  {journal} {\bibinfo  {journal} {Phys. Rev. B}\ }\textbf {\bibinfo {volume}
  {78}},\ \bibinfo {pages} {245103} (\bibinfo {year} {2008})}\BibitemShut
  {NoStop}%
\bibitem [{\citenamefont {Squires}(1996)}]{squires-book}%
  \BibitemOpen
  \bibfield  {author} {\bibinfo {author} {\bibfnamefont {G.~L.}\ \bibnamefont
  {Squires}},\ }\href@noop {} {\emph {\bibinfo {title} {Introduction to the
  Theory of Thermal Neutron Scattering}}}\ (\bibinfo  {publisher} {Dover
  Publications},\ \bibinfo {address} {Mineola, N.Y.},\ \bibinfo {year}
  {1996})\BibitemShut {NoStop}%
\bibitem [{\citenamefont {Walters}\ \emph {et~al.}(2009)\citenamefont
  {Walters}, \citenamefont {Perring}, \citenamefont {Caux}, \citenamefont
  {Savici}, \citenamefont {Gu}, \citenamefont {Lee}, \citenamefont {Ku},\ and\
  \citenamefont {Zaliznyak}}]{walters-natphys-2009}%
  \BibitemOpen
  \bibfield  {author} {\bibinfo {author} {\bibfnamefont {A.~C.}\ \bibnamefont
  {Walters}}, \bibinfo {author} {\bibfnamefont {T.~G.}\ \bibnamefont
  {Perring}}, \bibinfo {author} {\bibfnamefont {J.~S.}\ \bibnamefont {Caux}},
  \bibinfo {author} {\bibfnamefont {A.~T.}\ \bibnamefont {Savici}}, \bibinfo
  {author} {\bibfnamefont {G.~D.}\ \bibnamefont {Gu}}, \bibinfo {author}
  {\bibfnamefont {C.~C.}\ \bibnamefont {Lee}}, \bibinfo {author} {\bibfnamefont
  {W.}~\bibnamefont {Ku}}, \ and\ \bibinfo {author} {\bibfnamefont {I.~A.}\
  \bibnamefont {Zaliznyak}},\ }\href@noop {} {\bibfield  {journal} {\bibinfo
  {journal} {Nature Physics}\ }\textbf {\bibinfo {volume} {5}},\ \bibinfo
  {pages} {867} (\bibinfo {year} {2009})}\BibitemShut {NoStop}%
\bibitem [{\citenamefont {Coldea}\ \emph {et~al.}(2001)\citenamefont {Coldea},
  \citenamefont {Hayden}, \citenamefont {Aeppli}, \citenamefont {Perring},
  \citenamefont {Frost}, \citenamefont {Mason}, \citenamefont {Cheong},\ and\
  \citenamefont {Fisk}}]{coldea-prl-2001}%
  \BibitemOpen
  \bibfield  {author} {\bibinfo {author} {\bibfnamefont {R.}~\bibnamefont
  {Coldea}}, \bibinfo {author} {\bibfnamefont {S.~M.}\ \bibnamefont {Hayden}},
  \bibinfo {author} {\bibfnamefont {G.}~\bibnamefont {Aeppli}}, \bibinfo
  {author} {\bibfnamefont {T.~G.}\ \bibnamefont {Perring}}, \bibinfo {author}
  {\bibfnamefont {C.~D.}\ \bibnamefont {Frost}}, \bibinfo {author}
  {\bibfnamefont {T.~E.}\ \bibnamefont {Mason}}, \bibinfo {author}
  {\bibfnamefont {S.~W.}\ \bibnamefont {Cheong}}, \ and\ \bibinfo {author}
  {\bibfnamefont {Z.}~\bibnamefont {Fisk}},\ }\href@noop {} {\bibfield
  {journal} {\bibinfo  {journal} {Phys. Rev. Lett.}\ }\textbf {\bibinfo
  {volume} {86}},\ \bibinfo {pages} {5377} (\bibinfo {year}
  {2001})}\BibitemShut {NoStop}%
\bibitem [{\citenamefont {Headings}\ \emph {et~al.}()\citenamefont {Headings},
  \citenamefont {Hayden}, \citenamefont {Coldea},\ and\ \citenamefont
  {Perring}}]{headings-arxiv}%
  \BibitemOpen
  \bibfield  {author} {\bibinfo {author} {\bibfnamefont {N.~S.}\ \bibnamefont
  {Headings}}, \bibinfo {author} {\bibfnamefont {S.~M.}\ \bibnamefont
  {Hayden}}, \bibinfo {author} {\bibfnamefont {R.}~\bibnamefont {Coldea}}, \
  and\ \bibinfo {author} {\bibfnamefont {T.~G.}\ \bibnamefont {Perring}},\
  }\href@noop {} {\bibinfo  {journal} {arXiv:1009.2915}\ }\BibitemShut
  {NoStop}%
\bibitem [{\citenamefont {Kim}\ \emph {et~al.}(1999)\citenamefont {Kim},
  \citenamefont {Aharony}, \citenamefont {Birgeneau}, \citenamefont {Chou},
  \citenamefont {Entin-Wohlman}, \citenamefont {Erwin}, \citenamefont {Greven},
  \citenamefont {Harris}, \citenamefont {Kastner}, \citenamefont {Korenblit},
  \citenamefont {Lee},\ and\ \citenamefont {Shirane}}]{kim-prl-1999}%
  \BibitemOpen
\bibfield  {journal} {  }\bibfield  {author} {\bibinfo {author} {\bibfnamefont
  {Y.~J.}\ \bibnamefont {Kim}}, \bibinfo {author} {\bibfnamefont
  {A.}~\bibnamefont {Aharony}}, \bibinfo {author} {\bibfnamefont {R.~J.}\
  \bibnamefont {Birgeneau}}, \bibinfo {author} {\bibfnamefont {F.~C.}\
  \bibnamefont {Chou}}, \bibinfo {author} {\bibfnamefont {O.}~\bibnamefont
  {Entin-Wohlman}}, \bibinfo {author} {\bibfnamefont {R.~W.}\ \bibnamefont
  {Erwin}}, \bibinfo {author} {\bibfnamefont {M.}~\bibnamefont {Greven}},
  \bibinfo {author} {\bibfnamefont {A.~B.}\ \bibnamefont {Harris}}, \bibinfo
  {author} {\bibfnamefont {M.~A.}\ \bibnamefont {Kastner}}, \bibinfo {author}
  {\bibfnamefont {I.~Y.}\ \bibnamefont {Korenblit}}, \bibinfo {author}
  {\bibfnamefont {Y.~S.}\ \bibnamefont {Lee}}, \ and\ \bibinfo {author}
  {\bibfnamefont {G.}~\bibnamefont {Shirane}},\ }\href@noop {} {\bibfield
  {journal} {\bibinfo  {journal} {Phys. Rev. Lett.}\ }\textbf {\bibinfo
  {volume} {83}},\ \bibinfo {pages} {852} (\bibinfo {year} {1999})}\BibitemShut
  {NoStop}%
\bibitem [{\citenamefont {R{\o}nnow}\ \emph {et~al.}(2001)\citenamefont
  {R{\o}nnow}, \citenamefont {McMorrow}, \citenamefont {Coldea}, \citenamefont
  {Harrison}, \citenamefont {Youngson}, \citenamefont {Perring}, \citenamefont
  {Aeppli}, \citenamefont {Sylju{\aa}sen}, \citenamefont {Lefmann},\ and\
  \citenamefont {Rischel}}]{ronnow-prl-2001}%
  \BibitemOpen
  \bibfield  {author} {\bibinfo {author} {\bibfnamefont {H.~M.}\ \bibnamefont
  {R{\o}nnow}}, \bibinfo {author} {\bibfnamefont {D.~F.}\ \bibnamefont
  {McMorrow}}, \bibinfo {author} {\bibfnamefont {R.}~\bibnamefont {Coldea}},
  \bibinfo {author} {\bibfnamefont {A.}~\bibnamefont {Harrison}}, \bibinfo
  {author} {\bibfnamefont {I.~D.}\ \bibnamefont {Youngson}}, \bibinfo {author}
  {\bibfnamefont {T.~G.}\ \bibnamefont {Perring}}, \bibinfo {author}
  {\bibfnamefont {G.}~\bibnamefont {Aeppli}}, \bibinfo {author} {\bibfnamefont
  {O.}~\bibnamefont {Sylju{\aa}sen}}, \bibinfo {author} {\bibfnamefont
  {K.}~\bibnamefont {Lefmann}}, \ and\ \bibinfo {author} {\bibfnamefont
  {C.}~\bibnamefont {Rischel}},\ }\href@noop {} {\bibfield  {journal} {\bibinfo
   {journal} {Phys. Rev. Lett.}\ }\textbf {\bibinfo {volume} {87}},\ \bibinfo
  {pages} {037202} (\bibinfo {year} {2001})}\BibitemShut {NoStop}%
\bibitem [{\citenamefont {Christensen}\ \emph {et~al.}(2007)\citenamefont
  {Christensen}, \citenamefont {R{\o}nnow}, \citenamefont {McMorrow},
  \citenamefont {Harrison}, \citenamefont {Perring}, \citenamefont {Enderle},
  \citenamefont {Coldea}, \citenamefont {Regnault},\ and\ \citenamefont
  {Aeppli}}]{christensen-pnas-2007}%
  \BibitemOpen
  \bibfield  {author} {\bibinfo {author} {\bibfnamefont {N.~B.}\ \bibnamefont
  {Christensen}}, \bibinfo {author} {\bibfnamefont {H.~M.}\ \bibnamefont
  {R{\o}nnow}}, \bibinfo {author} {\bibfnamefont {D.~F.}\ \bibnamefont
  {McMorrow}}, \bibinfo {author} {\bibfnamefont {A.}~\bibnamefont {Harrison}},
  \bibinfo {author} {\bibfnamefont {T.~G.}\ \bibnamefont {Perring}}, \bibinfo
  {author} {\bibfnamefont {M.}~\bibnamefont {Enderle}}, \bibinfo {author}
  {\bibfnamefont {R.}~\bibnamefont {Coldea}}, \bibinfo {author} {\bibfnamefont
  {L.~P.}\ \bibnamefont {Regnault}}, \ and\ \bibinfo {author} {\bibfnamefont
  {G.}~\bibnamefont {Aeppli}},\ }\href@noop {} {\bibfield  {journal} {\bibinfo
  {journal} {Proc. Natl. Acad. Sci. U.S.A.}\ }\textbf {\bibinfo {volume}
  {104}},\ \bibinfo {pages} {15264} (\bibinfo {year} {2007})}\BibitemShut
  {NoStop}%
\bibitem [{\citenamefont {Huberman}\ \emph {et~al.}(2005)\citenamefont
  {Huberman}, \citenamefont {Coldea}, \citenamefont {Cowley}, \citenamefont
  {Tennant}, \citenamefont {Leheny}, \citenamefont {Christianson},\ and\
  \citenamefont {Frost}}]{huberman-prb-2005}%
  \BibitemOpen
  \bibfield  {author} {\bibinfo {author} {\bibfnamefont {T.}~\bibnamefont
  {Huberman}}, \bibinfo {author} {\bibfnamefont {R.}~\bibnamefont {Coldea}},
  \bibinfo {author} {\bibfnamefont {R.~A.}\ \bibnamefont {Cowley}}, \bibinfo
  {author} {\bibfnamefont {D.~A.}\ \bibnamefont {Tennant}}, \bibinfo {author}
  {\bibfnamefont {R.~L.}\ \bibnamefont {Leheny}}, \bibinfo {author}
  {\bibfnamefont {R.~J.}\ \bibnamefont {Christianson}}, \ and\ \bibinfo
  {author} {\bibfnamefont {C.~D.}\ \bibnamefont {Frost}},\ }\href@noop {}
  {\bibfield  {journal} {\bibinfo  {journal} {Phys. Rev. B}\ }\textbf {\bibinfo
  {volume} {72}},\ \bibinfo {pages} {014413} (\bibinfo {year}
  {2005})}\BibitemShut {NoStop}%
\end{thebibliography}%

\end{document}